\documentclass[letterpaper]{article}
\usepackage{aaai}
\usepackage{times}
\usepackage{helvet}
\usepackage{courier}
\usepackage{url}
\usepackage{graphicx}
\usepackage{natbib}
\usepackage{booktabs}
\usepackage{caption}
\usepackage{tabularx}
\usepackage{xspace}
\newcommand{\sys}{QuizRank\xspace}
\usepackage{listings}
\usepackage{enumitem}
\setlist[itemize]{align=parleft,left=0pt..1em,noitemsep, nolistsep}

\usepackage{xcolor}

\begin{document}

\title{\sys: Picking Images by Quizzing VLMs}
\author{Tenghao Ji \& Eytan Adar\\University of Michigan}

\maketitle
\begin{abstract}
Images play a vital role in improving the readability and comprehension of Wikipedia articles by serving as `illustrative aids.' However, not all images are equally effective and not all Wikipedia editors are trained in their selection. We propose \sys, a novel method of image selection that leverages large language models (LLMs) and vision language models (VLMs) to rank images as learning interventions. Our approach transforms textual descriptions of the article's subject into multiple-choice questions about important visual characteristics of the concept. We utilize these questions to quiz the VLM: the better an image can help answer questions, the higher it is ranked. To further improve discrimination between visually similar items, we introduce a \textit{Contrastive \sys} that leverages differences in the features of target (e.g., a \textit{Western Bluebird}) and distractor concepts (e.g., \textit{Mountain Bluebird}) to generate questions. We demonstrate the potential of VLMs as effective visual evaluators by showing a high congruence with human quiz-takers and an effective discriminative ranking of images.
\end{abstract}

\section{Introduction}
Wikipedia guides editors to select ``images [that] look \textit{like} what they are meant to illustrate'' (emphasis in original) and to serve as an ``illustrative aid''~\citep{wikimg}. Images are important in improving the comprehension of textual content by a reader~\citep{clark2010graphics} and numerous studies show that well-integrated visuals can improve comprehension, retention, and engagement \citep{guo2020you, harp1997role, carney2002pictorial}. However, selecting these images is challenging. As of September 2025 there are over 7 million article pages in the English Wikipedia and 127 million images on the Wikipedia commons. Editors may not know what makes a good image or will need to make trade-offs in their choices (e.g., should we show a bird in context? in all its variations?). The result may be that, for many pages, images can be generic, unclear, misaligned, or poorly matched to the content of the article, thus limiting their communicative effectiveness~\citep{rama2022large, navarrete2020image}.

There are natural opportunities to help editors by automatically identifying good images based on a concept's textual description\footnote{A \textit{concept} is often the subject of a Wikipedia article, but we allow for an article to talk about multiple \textit{concepts}, each may have their own image}. Existing efforts have focused mainly on similarity metrics between text and image captions~\citep{agrawal2011enriching,aletras2013representing,leake2020generating,singh2023enhancing} or isolating representative examples from an image set~\citep{han2020extracting}. Other approaches have tried to directly use the LLM/VLM as a judge~\citep{li2024generation} by prompting the model to evaluate the image's fit to a section (e.g.,~\citep{yang2024toward}). However, these approaches can be somewhat coarse and misaligned from the actual intent of the image--to serve as an educational aid. To address this, we propose \sys, a system that directly uses learning assessments to rank possible images. Quizzes have two main functions. The first is to assess an individual student. The second, which is key for our approach, is that assessments serve as programmatic evaluations. That is, how effective was the teacher/module/lesson, or in our case \textit{the image}, in improving learning?

\begin{figure*}
    \centering
    \includegraphics[width=\linewidth]{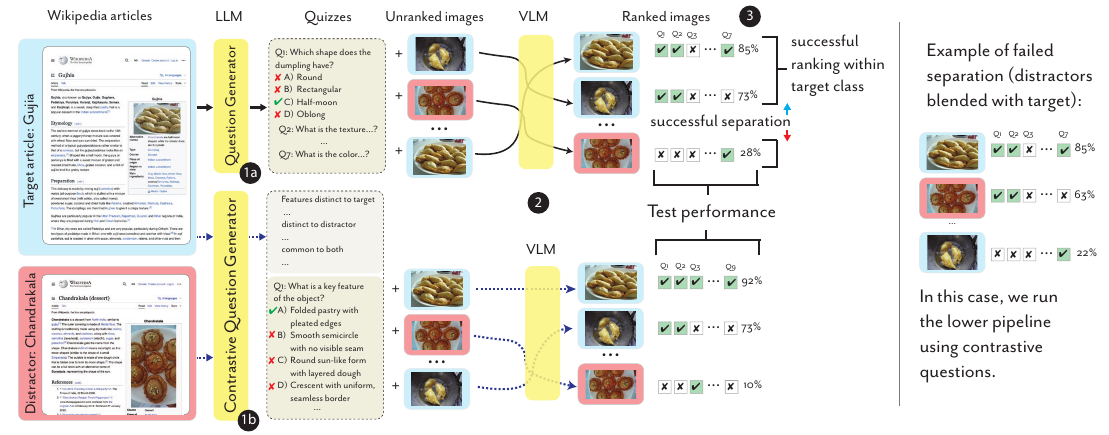}
    \caption{Example of the QuizRank algorithm being applied with both standard (top) and contrastive ranking (bottom). In both situations a `quiz' is generated based on the description of the target option (and, in the case of contrastive, one or more distractors). The VLM will attempt to answer the questions for each of these quizzes per image allowing us to rank them. Note: the example questions and images here are only illustrative, but a complete run and analysis for the Guija/Chandrakala case is available in the Appendix.}
    \label{fig:algorithm}
\end{figure*}

\sys (graphically depicted in Figure~\ref{fig:algorithm}) analyzes the text of the article to identify the visual characteristics of the concept (e.g., ``the statue is bronze''). From these, the system constructs a set of multiple choice questions (e.g., ``what is the statue made from? A) Bronze, B) Marble, etc.''). By coupling the images and this generated exam. \sys uses a vision-language model (VLM) to score each possible image. The intuition is that the more helpful the image is for the VLM in taking the quiz, the better the image. 

To select better images, \sys can generate quizzes on both the original concept and through an analysis of reasonable distractors (\textit{contrastive question generation}). For example, in Figure~\ref{fig:algorithm} we have two possible images for a target concept---Gujia, a kind of fried Indian pastry---and a third image for a Chandrakala (a different fried pastry). A good image would be a perfect `representation' of the target concept, while at the same time highlighting the distinguishing features of the concept relative to similar concepts.  Our contrastive setup---which draws inspiration from past work on visual disambiguation~\citep{plantguide, bruce1994recognizing}---encourages the model to focus on specific differences between similar entities and improves the robustness of image selection.

To evaluate \sys, we focus on the correspondence between the VLM's performance with a human's on answering questions given different images. We show that the VLM is roughly analogous to the human performance in both scoring images and distinguishing in-class vs. out-of-class images. We demonstrate the effectiveness of the approach in selecting images on a wide array of Wikipedia concepts.

\section{Related Work}

Our work builds on previous research on question generation, multimodal learning, visual comprehension in educational contexts, and Wikipedia-based image evaluation. 

\paragraph{Question Generation}
\sys directly builds on work in question generation (QG). Original efforts in the domain used highly templatized and domain-specific approaches (e.g. ~\citep{kurdi2020systematic}). However, with the prevalence of LLMs, more modern QG systems provide more sophisticated analysis of text for question generation (e.g., ~\citep{richards2023readerquizzer, shimmei2023machine, moore2023assessing, lu2024generativestudents}). \sys uses a similar prompting approach to construct questions, but focuses on visual properties. Additionally, we develop an approach for generating contrastive questions that can help distinguish both misclassified and out-of-class objects by emphasizing the visual properties of the concept.

Multiple choice questions (MCQs) consist of a question stem, a correct answer, and distractors. Although simple a the high level, these still need to be well designed to ensure proper assessment~\citep{https://doi.org/10.1111/j.1745-3992.2005.00006.x}. In building \sys, we apply established guidelines and good practices (e.g.,~\citep{haladyna2004developing, little2012multiple, schwartz2016abcs}) to create our MCQs.

\paragraph{Images in Wikipedia and Multilingual Visual Knowledge.}
Psychological studies have demonstrated the positive effect of illustrations on comprehension and learning (e.g.,~\citep{carney2002pictorial,guo2020you}. Because of this, dictionaries and encyclopedias include illustrations and images for many specific purposes~\citep{stein1991illustrations}. Similarly, there are many benefits to integrating images into a Wikipedia article~\citep{wikimg}. However, past work on Wikipedia images highlighted the challenges of image selection~\citep{rama2022large}. This reinforces the need for assessments for image appropriateness and educational value.

Silva et al.~\citep{silva2024imagine} constructed a benchmark of 94 Wikipedia concepts with 470 multiple-choice questions to test: general knowledge (e.g., \textit{when was the tower built?}); visual feature recall (e.g., \textit{what color is the Western Bluebird's chest?}); and concept recognition (e.g., \textit{which of these images represents a Mountain Bluebird?}). Their study, which tested human participants, found that the utility of the image in Wikipedia is highly dependent on the semantic alignment between the image and the content of the article. We use the same benchmark, but shift focus to evaluating image relevance through vision-language models (VLMs) rather than human annotators. 

\paragraph{Multi-modal Representation and Vision-Language Alignment.}
There are a number of datasets that provide image-to-text alignment. For example, Srinivasan et al.~\citep{srinivasan2021wit} introduced a multilingual Wikipedia-based image–text corpus that provides grounding for training and evaluating VLMs. Our work builds on efforts to link text to images, but focuses on learning objectives centered on the visual properties of the objects.

Prior work has demonstrated that vision-language pretraining using natural language supervision enables transfer to many downstream tasks~\citep{radford2021learning}. We leverage this idea to design visual comprehension questions that can differentiate between semantically appropriate and `distractor' images.
Other efforts categorized different types of image–text relations, offering taxonomies for alignment assessment~\citep{gyselinck1999role,Marsh2003:0022-0418:647}.  In contrast, our method evaluates alignment functionally through QA performance. Finally, Wikimedia's own Add-an-Image initiative~\citep{wikimedia2022addimage} explores crowd-based image suggestion tasks. Our framework offers a scalable alternative using VLMs.

Together, these studies motivate our approach: to move beyond subjective human assessments by building an objective, interpretable framework that measures an image's semantic and visual relevance.

\section{\sys Design}
\sys utilizes a multi-stage method to evaluate and rank image quality as illustrative aids. A central goal of the algorithm is high-quality illustrative images. We define high quality as those images that are good \textit{in-class} and \textit{between-class} representatives.

\textbf{In-Class Representation}: Good images will illustrate key visual features of the target concept. For example, if it is important that a statue is made of bronze, is 9 feet tall, and is missing an arm, the image should show this. To operationalize this feature, we focus our generated questions so that the VLM will validate if the image contains the feature. In other words, the better the image, the more help it will provide to the VLM in answering the questions.

\textbf{Between-Class Representation}: Good images will also highlight the specific properties of the concept that \textit{are distinct from related concepts}. For example, if a person were to learn what a Western Bluebird looked like from the image, would they later be able to distinguish a Western Bluebird they see in the wild from a Mountain Bluebird? We operationalize this by creating questions for the VLM to answer that when given in-class images, the VLM will be more likely to answer correctly than seeing distractor images (i.e., good discriminative questions). 

\sys achieves this through four steps (see Figure~\ref{fig:algorithm}). First, we generate MCQs from the article's content (1a) and/or MCQs that also take into account distinctions with related articles--a contrastive refinement step (1b). For each image we are considering, we feed this quiz to a VLM (2) to see how many it can answer based on the image. Finally, based on how well each image performs to `help' the VLM answer the questions, we can rank the images (3).

\subsection{Base Question Generation}
\sys leverages past work on generating multiple choice questions using LLMs. The input to this process is the concept's Wikipedia page. This can be optionally focused to those sub-sections that are more likely to contain visual properties of the concept (e.g., Architecture, Appearance, Description, Characteristics, Design, Details, etc.). The LLM prompt (see Appendix) was designed to utilize `good' multiple choice design approaches~\citep{haladyna2004developing}. \sys's prompt uses chain-of-thought style prompting with an emphasis on targeting those features that might be inferred visually from an image~\citep{wei2022chain}. To achieve this, it emphasizes that features such as texture, material, and structure should be prioritized. To reduce the likelihood that the question will `leak' information about the concept itself, the prompt tries to enforce generalizations. For example, for Donatello's statue of David\footnote{\url{https://w.wiki/BGsp}}, \sys generated the questions: ``What material is the statue made of?'' (with \textit{Bronze}, \textit{Marble}, \textit{Iron}, and \textit{Wood} as possible answers); ``What is the male figure wearing?'' (the article mentions the hat); and ``What notable expression does the figure have?'' This type of generalization reduces the ability of the model to make inferences (i.e., the LLM may be aware that the Donatello's David is made of Bronze) and better simulates the information a human might use in answering the questions. 

We do not enforce a minimum or maximum number of questions as this is often correlated with the number of visual features described in the article. Short articles provide fewer visual descriptions than long ones. In our data (see below) we find that \sys produces a minimum of 4 questions and a maximum of 11 (with a mode of 5). This number naturally depends on the length of the Wikipedia article's characterization of the visual properties.

\subsection{Contrastive refinement}
An encyclopedia is not intended as a field guide (e.g., to help birders identify and distinguish between birds~\citep{birdguide}). However, images as illustrative aids should not only serve as good exemplars of the target class, but should also emphasize the key distinctive qualities of that class in relation to others. For example, we can consider images of the Caiman\footnote{\url{https://w.wiki/5Cvt}}, a type of alligator that has interesting skin and teeth. We can look at an image and ask questions about its skin and teeth to see if that image is a good representative. However, because the Caiman shares properties with other related animals (e.g., the Chinese Alligator), looking at images of non-Caiman alligators may also generate the correct result. In situations where we want to find images that are good instructional aids in knowing what a Caiman looks like, these questions might not help. 

To better support between-class effectiveness, \sys implements a contrastive refinement step (Figure~\ref{fig:algorithm}-1b). Specifically, we utilize a prompt that can take both the visual properties of the target concept and one or more similar concepts to first identify the distinct and overlapping features of these concepts (see the Appendix for the complete prompt). From the \textit{unique} visual characteristics, we are able to generate a more specific set of questions. Note that we do not \textit{always} need to generate contrastive questions. In some situations there are simply no similar concepts, and the VLM is unlikely to answer correctly given out-of-class images. In our implementation, we test these out-of-class images (e.g., of the Chinese Alligator) on the base questions. Only in those situations where the VLM answers the class-specific questions (e.g., about the Caiman) \textit{correctly} with the out-of-class images do we generate contrastive questions. In Wikipedia, related contrastive concepts are selected from related linked articles or from other concepts/topics in the same Wikipedia class (e.g., \textit{Crocodilians by location}). Note that if we were to generate all possible questions about the visual properties of a concept, this approach may not be as valuable. However, creating contrastive questions may be a more cost-effective approach in that we can focus on only important questions. 

\subsection{Quizzing the VLM}
To determine if an image is good, we can submit the image to the VLM with our MCQs to see if the image helps the VLM answer the questions (prompt in the Appendix). The more questions the VLM correctly answers, the more confident we can be that the image is good.

To ensure that the VLM is using the image as evidence, we attempt to prevent the model from utilizing information other than what is in the image. As described before, we take steps to avoid leaking the concept's identity by never explicitly mentioning it in the questions. In addition, the prompt strictly instructs the model to rely solely on visible content in the image, ignoring prior knowledge, indexed facts, or textual context. As a final check, the prompt also tries to force the VLM to generate reasoning that describes why it chose the answer given what it saw in the image. For example, in response to the question about the statue's material, the model reasoned:

\begin{lstlisting}[basicstyle=\footnotesize]
To determine the material of the statue based on the image, I'll analyze the visual characteristics:

1. **Color and Texture**: The statue has a dark, glossy appearance that resembles metal. This sheen is typical of metals like bronze.

2. **Detailing**: The fine detailing in the hair and the facial features suggest a material that can hold intricate designs well, like bronze or marble. However, the color leans more towards bronze.

3. **Overall Visual**: The statue doesn't show the grain typical of wood, the light color, and translucence of marble, nor does it have the heavier, less refined appearance of iron.

Based on these observations, the most likely material is A) Bronze.
Correct Answer: A) Bronze
\end{lstlisting}

The prompt also allows for the answer, ``I can’t answer that based on the image.'' 

\subsection{Answer-based image ranking}
A final step (Figure~\ref{fig:algorithm}-4) calculates the overall quiz score to rank the images as a form of model-as-judge~\citep{li2024generation}. In the simplest instance, we can calculate the normalized number of correct answers for each image and rank based on this. That is, how many questions did each image help the VLM answer? However, there are situations where no \textit{one} image can answer all the questions correctly. \sys can keep individual scores (e.g., a matrix of images versus questions) to identify \textit{bundles} of images that may be useful. Other types of scoring (e.g., clustering questions based on correlated answers, weighting by question difficulty, or the importance of different properties) may also be possible but are left for future work.

\section{Evaluation}
To evaluate our approach, we focus on two high-level objectives. First, we would like to validate the \textit{correspondence} of the VLM with human perception. Ideally, the VLM will answer the questions as a human would, using only information from the image. If the VLM benefits from the instructional aid (the image), we can infer that a human may also benefit.

\begin{itemize}
\item \textbf{RQ1}:  Do humans and VLMs answer questions in similar ways when observing the image?
\item \textbf{RQ2}:  Does the VLM rely on external information (e.g., training text) when answering?
\end{itemize}

Second, we validate the \textit{effectiveness} of the approach in isolating better images on Wikipedia.

\begin{itemize}
    \item \textbf{RQ3}: Can \sys identify better images than Wikipedia's crowdsourced approach?
    \item \textbf{RQ4}: How much variation is there in the quality of available in-class images on Wikipedia?
\end{itemize}
 
Finally, we end with a brief analysis of the stability of \sys results under different quiz sizes.

\subsection{Data and Methods}
We conducted two primary experiments to answer our research questions. We started with the 94 concepts selected in~\citep{silva2024imagine} (we refer the interested reader there for full details). Briefly, the concepts were chosen to be from popular, but less known Wikipedia pages. Concepts were selected from a wider range of Wikipedia categories: biology, architecture, food and drink, and art.  From the original 94 concepts, we excluded five because they either focused on non-visual subjects (e.g., biographical articles) or were too short to generate meaningful visual questions. The Wikipedia articles for the 89 concepts were submitted to \sys for question generation, initially using the non-contrastive pipeline. We used OpenAI's GPT-4o as an LLM model (with additional validation using LLaMA-4-Scout-17B) to generate the per-concept quizzes.

For each concept, we curated four images: three images depicting the target concept (selected from the Wikipedia Commons) and one image of a visually similar but incorrect object (a distractor). The distractor was chosen by identifying concepts within the broad class (e.g., the Chinese Alligator for the Caiman). To avoid selecting obviously bad images, we required that an image be used on a Wikipedia article page (any language). Each image and the associated quizzes were submitted to GPT-4o for the VLM step of \sys (again validating with LLaMA). If the distractor image performed similarly to the correct images (i.e., difference in score $<$ 2), we triggered the Contrastive \sys approach to create new quizzes and a new grade. 

We collected human quiz scores on a subset of 40 concepts.  We recruited 320 participants through the Prolific platform and administered the task via a Qualtrics survey (payment for this task was \$1-\$1.25). Each participant was randomly assigned five concepts. For each concept, participants were shown one randomly selected image for the concept and asked to answer the questions generated by \sys. We aimed for each image to be evaluated by two different participants. This required 320 participants (40 concepts $\times$ 4 images $\times$ 2 labels). As with the VLM experiments, our goal was to understand: did quiz performance improve with the presence of the image? 

\subsection{VLM-Human Correspondence}
\begin{figure}[ht]
    \centering
    \includegraphics[width=0.4\textwidth]{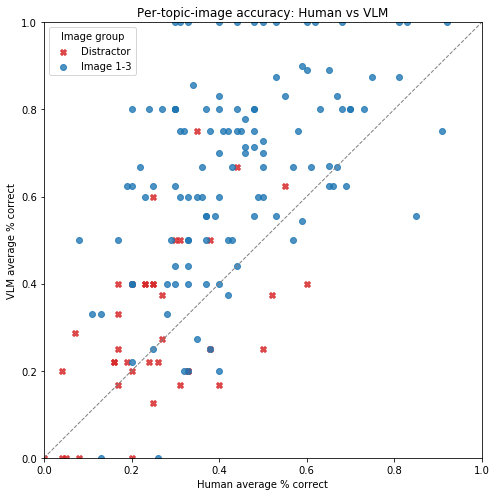} 
    \caption{Per-image average accuracy comparison between human participants and the VLM. Red points represent distractor images. The dotted line indicates parity.}
    \label{fig:human_vlm_alignment}
\end{figure}

We compared the average accuracy per image from both VLM and human sources across the 40 concepts (see Figure~\ref{fig:human_vlm_alignment}). Each point represents a single image: blue circles denote images illustrating the correct concept (Images 1--3), while red asterisks denote distractor images.

We observe a strong correlation (Pearson $r=0.6125$, $p<.001$). This supports the idea that the images that were the most effective in answering the questions helped both humans and VLMs in the same way (\textbf{RQ1}). We also note that the distractors were generally located in the lower left quadrant, indicating that the baseline questions had some discriminative power. In general, the answers produced by the VLM tended to be higher than those of humans. This may be due to individual differences (e.g., cultural or concept familiarity/interest) and different levels of attention. It may also reflect a stronger inferential ability for the VLM (i.e., it can better reason through the choices). In practical use of \sys, this may not be a critical factor. Although the baseline score (the floor) may be higher, all images will be evaluated by the VLM. Thus, it is unlikely that any specific image will have some advantage.

\subsubsection{Open-source models}
We re-tested the full \sys{} pipeline using the open-source LLaMA-4-Scout-17B model. The setup mirrored our primary experiment on 40 concepts (4 images per concept, with 3 in-class and 1 distractor image), allowing direct comparison.

Our results show strong alignment between LLaMA and both the human and GPT-4o judgments. The Pearson correlation between LLaMA's image rankings and those of human participants was $r = 0.613$ ($p = 7.51 \times 10^{-18}$), nearly identical to what we observed with GPT-4o ($r = 0.6125$). In terms of agreement with GPT-4o rankings, we computed the per-concept Spearman correlation and found that: 28 concepts had very strong correlations ($\rho = 0.8$–$1.0$); 4 concepts were strong ($\rho = 0.6$–$0.79$); 2 concepts were moderate ($\rho = 0.4$–$0.59$); 3 each were weak ($\rho = 0.2$–$0.39$) or very weak ($\rho = 0$–$0.19$). Overall, 32 out of 40 concepts (80\%) showed strong or very strong agreement with GPT-4o, and 26 concepts (65\%) had perfect rank correlation. In general, these findings reinforce the consistency and robustness of the \sys{} method across different models and the results are likely to improve with newer models.

\subsubsection{Between Concept Separation}
\begin{figure}[t]
\centering
\begin{minipage}[t]{0.42\textwidth}
    \centering
    \includegraphics[width=\textwidth]{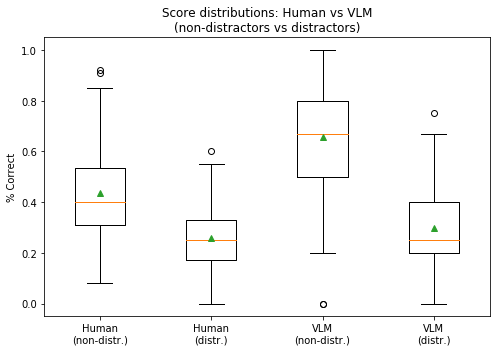}
    \caption{Score distributions for human and VLM across distractor and non-distractor images. Green triangles denote means.}
    \label{fig:score_dist}
\end{minipage}
\hfill
\begin{minipage}[t]{0.48\textwidth}
    \centering
    \scriptsize
    \setlength{\tabcolsep}{2pt}  %
    \begin{tabular}{lcccccc}
    \toprule
    \textbf{Comparison} & \textbf{n1} & \textbf{n2} & \textbf{KW\_stat} & \textbf{KW\_p} & \textbf{ANOVA\_F} & \textbf{ANOVA\_p} \\
    \midrule
    Human non vs distr & 120 & 40 & 30.420& 0.000 & 33.717 & 0.0000 \\
    VLM non vs distr   & 120 & 40 & 53.331 & 0.000 & 83.027 & 0.0000 \\
    Human vs VLM (non) & 120 & 120 & 57.726 & 0.000 & 70.836 & 0.0000 \\
    Human vs VLM (distr) & 40 & 40 & 0.753 & 0.3857 & 1.155 & 0.2858 \\
    \bottomrule
    \end{tabular}
    \captionof{table}{Statistical test results comparing human and VLM performance on different image types.}
    \label{tab:stat_test_summary}
\end{minipage}
\end{figure}

Figure~\ref{fig:score_dist} provides an aggregate view of the score distributions for human and VLM responses across distractor and non-distractor images. Human participants performed significantly better on non-distractor images ($M=0.44$, $SD=0.18$) than on distractors ($M=0.26$, $SD=0.14$). This difference is statistically significant according to both Kruskal–Wallis ($\chi^2=30.42$, $p<.001$) and ANOVA ($F=33.72$, $p<.001$) tests. Similarly, the VLM also scored significantly higher on non-distractor images ($M=0.66$, $SD=0.23$) than on distractors ($M=0.30$, $SD=0.19$), with both statistical tests confirming the distinction (Kruskal–Wallis: $\chi^2=53.33$, $p<.001$; ANOVA: $F=83.03$, $p<.001$). 

When directly comparing human and VLM performance on non-distractor images, the VLM significantly outperformed humans (Kruskal–Wallis: $\chi^2=57.73$, $p<.001$; ANOVA: $F=70.84$, $p<.001$). On distractor images, the VLM still scored marginally higher, but the difference was smaller and not significant (Kruskal–Wallis: $\chi^2=0.75$, $p=0.3857$; ANOVA: $F=1.155$, $p=0.286$). Anecdotally, the open-source model made a few more mistakes in producing a ranking that could separate the distractor, but these were limited.

\subsubsection{VLM Memorization}
A key concern for any generic LLM or VLM is that certain knowledge will become `baked in, and the model will have memorized the answer based on the text or because it has learned the caption for a certain image. For example, the \textit{Gayer-Anderson cat} is an Egyptian statue made of bronze with gold ornaments. This is a distinguishing characteristic of the stature that is captured in numerous text sources. If the VLM or LLM previously learned this information, the model may be able to correctly answer the questions regardless of the image content\footnote{In fact, for this example, GPT-4o does know the answer}. As described above, we endeavored to ensure that the prompts to not `leak' certain information to the model, and we force reasoning on the image rather than prompt/quiz text. In developing \sys, we did not identify cases in which the VLM used information outside of what the image provided in its reasoning.  In reality, we are less concerned about hints that the VLM can infer from the \textit{question} (i.e., if the quiz hints at an answer) as it will raise the score floor for all images, but the ranks will likely remain the same. 

However, if the image has been indexed, the VLM may be able to identify it and answer questions based on that identification. As a quick validation that the model's performance is not due to prior exposure to the test images (e.g., memorization or internal indexing), we conducted a controlled experiment. We used the eight concepts in our dataset that satisfied the constraint of having two images: one \textit{old} image that was uploaded to Wikimedia Commons before August 1, 2019, and one \textit{new} image uploaded after August 1, 2024. For this experiment we used the model \texttt{gpt-4o-2024-11-20}, which was trained on data up to October 2023. Thus, any image uploaded after August 2024 is unlikely to have been part of the model's training set. In other words, images uploaded after the latter threshold were not seen during the training of the VLM (i.e., unindexed), whereas those before the former threshold were likely part of the model's training data.

Both images were evaluated using the same set of questions generated from the article text. The results indicate no significant difference between old (likely indexed) and new (likely unindexed) images. Specifically, in 4 out of 8 concepts, both images led to the same number of correct answers; in 1 case, the indexed image outperformed the new one by one question; and in 3 cases, the unindexed image scored higher by one question. Image pairs were selected to convey comparable visual content and quality. This minimizes the risk that differences in question-answering performance are due to image clarity, composition, or informativeness, rather than indexing status (\textbf{RQ2}).

\subsection{In-Class Representation Quality}
By comparing to human performance, we have evidence that \sys can identify images that may be better for instructional aids. However, it is a question of whether this is actually needed in Wikipedia: does the underlying `wikiwork' already lead to the selection of good images?

To test this, we looked at the popularity of images (i.e., how many different pages an image is used in). If editors are effective, popularity might imply high quality. However, one language's editor may copy images from another (usually English~\citep{he2018the_tower_of_babel}) which may lead less effective choices to stick around and become popular.

We conducted a focused experiment on 13 random Wikipedia concepts (of the 89). For each concept, we selected 20 candidate images from Wikimedia Commons and recorded their frequency of usage---that is, the number of Wikipedia pages on which the image appears (for each of the 13 $\times$ 20 = 260 images). We computed a popularity score for each image as $\log_{10}(\mathrm{usage\ count} + 1)$ to account for skew and zero values. Each image/quiz pair was independently evaluated by the VLM. For each image, we computed the percentage of correctly answered questions and then standardized these scores within each concept using z-scores. This normalization allowed comparison across concepts with differing question counts and difficulty levels.

\begin{figure}[t]
    \centering
    \includegraphics[width=0.4\textwidth]{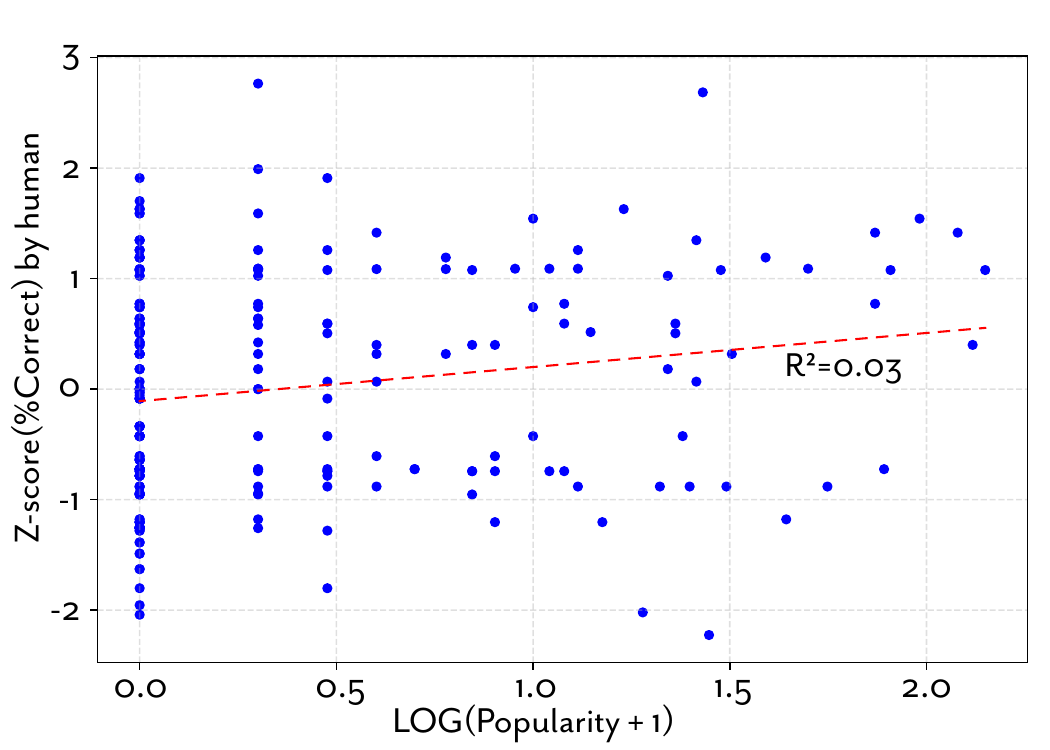}
    \caption{Correlation between image popularity (log scale) and standardized model performance (z-score of percent correct) across 13 Wikipedia concepts.}
    \label{fig:popularity_corr}
\end{figure}

Figure~\ref{fig:popularity_corr} shows a plot of image popularity ($x$-axis) against standardized model performance ($y$-axis). We observe a very weak positive trend, suggesting a weak correlation between an image's popularity and \sys score. However, there are clearly images that, for the purpose of serving as an instructional aid, may be better than the ones that are used (\textbf{RQ3}). This is not to say that there are not good reasons for selecting `sub-optimal' images. This may be for aesthetic reasons (a better composed or less blurry picture), or because the editor may be optimizing for some subset of visual properties. However, we believe that presenting the \sys rankings to an editor may be an opportunity to help them select better images. 

\subsection{Contrastive Pipeline Evaluation}
Figure~\ref{fig:human_vlm_alignment} partially demonstrates the ability of \sys to distinguish in-class images from distractors. We further analyze this based on the use of the contrastive question mechanism. For each of the 89 concepts we used one in-class image (the one used for the concept on the English language page) and an out-of-class distractor. Across the 89 concepts, we first evaluated model responses using the original basic (i.e., non-contrastive) question sets. In 57 concepts, the correct image outperformed the distractor image based on the VLM answers. In 26 concepts, the original and distractor images achieved equal scores. In only 6 cases did the distractor image outperform the in-class image. However, in 51 concepts, the difference in the number of correctly answered questions between in-class and distractor was 2 or less.

For concepts requiring further disambiguation (the 51 concepts), we applied the contrastive question generation mechanism. Among the 51 concepts that qualified, the original image performed better in 34 cases, performed equally in 16 cases, and was outperformed in only one case. These results suggest that our question-based image evaluation framework is generally robust and that the contrastive prompting strategy successfully improves discrimination when objects are visually similar.

\subsection{Within-Concept Variance}
\begin{figure}[t]
\centering
\includegraphics[width=0.8\linewidth]{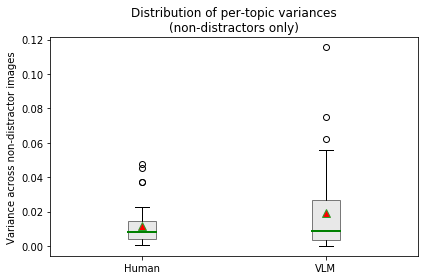}
\caption{Box plot of per-concept score variance across non-distractor images. Both Human and VLM show non-zero median variance, indicating discriminability among visually relevant candidates.}
\label{fig:variance_boxplot}
\end{figure}

One thing we would like to know is that the questions we generate actually have variance in score (we do not want all in-class images to score the same). As shown in Figure~\ref{fig:variance_boxplot}, the distribution of per-concept variances reveals that both humans and the VLM assign different scores to images within the same concept. This suggests that the generated visual questions are sufficiently discriminative for distractor removal (e.g., misclassified or bad images) but also fine-grained ranking among plausible images.

This finding aligns with prior work on visual image-text alignment in Wikipedia, which emphasizes that subtle visual differences can significantly affect comprehension and user perception~\citep{silva2024imagine, srinivasan2021wit}. Our results indicate that vision-language models, when guided by well-designed prompts, can reflect these distinctions through answer-based evaluation.

\subsection{Quiz Size}
It is difficult to know exactly how many questions are optimal for a quiz. This number may depend on the key visual characteristics described in the article (recall that shorter articles will yield fewer questions). Additionally, while we discouraged the LLM from generating similar questions the reality is that providing questions that are small variants of each other can better assess true understanding over random guessing. The `optimal' number of questions is hard to know and likely varies depending on the concept.

\begin{figure}[ht]
    \centering
    \includegraphics[width=0.48\textwidth]{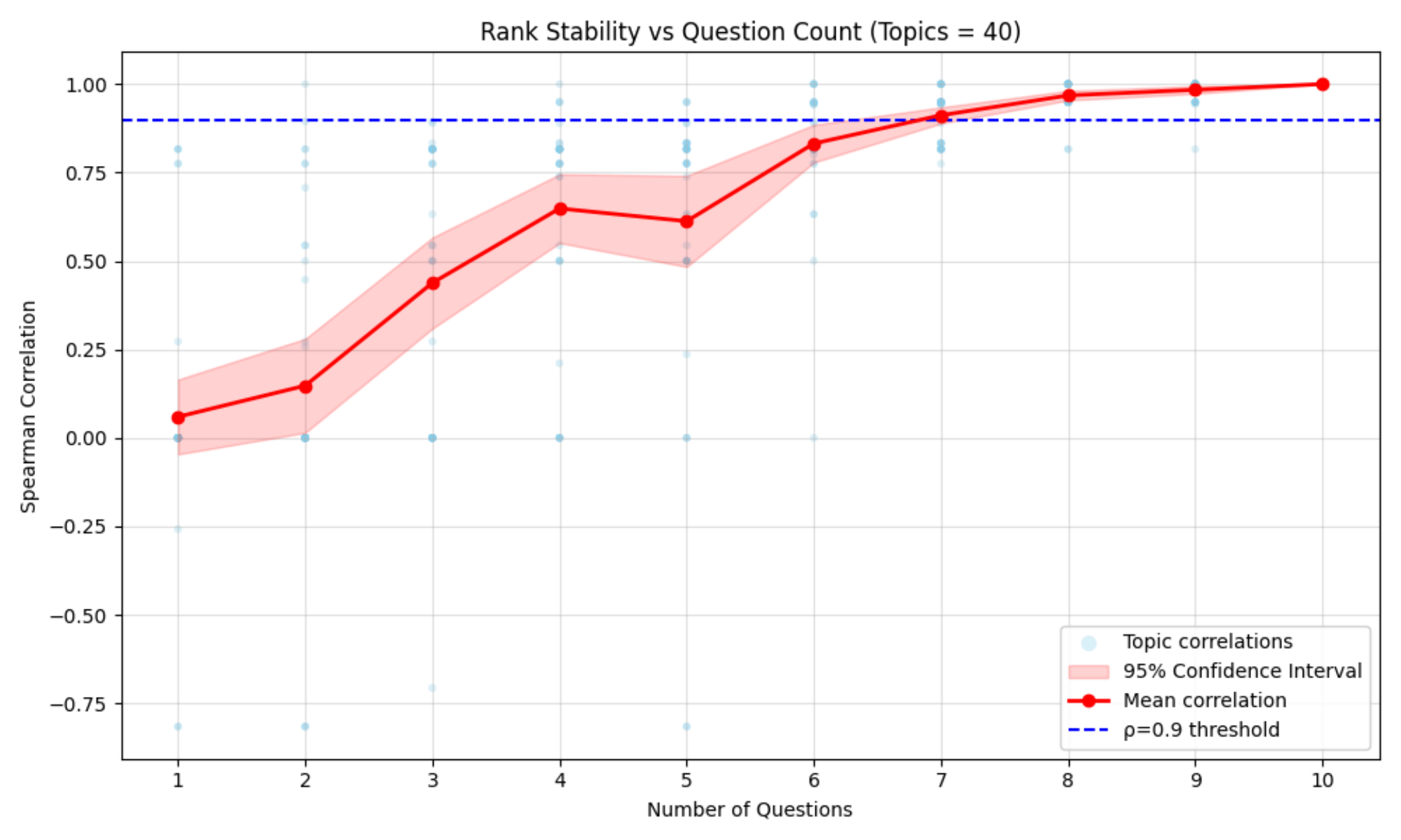}  
    \caption{Rank stability of concepts relative to a 10-question quiz.}
    \label{fig:asympt}
\end{figure}

Nevertheless, we were curious if the rank stability of images changed significantly based on the number of questions. To test this, we started with 10 questions per concept (which we anecdotally found to offer coverage of visual concepts over most articles in our test set). We randomly removed questions from this to produce quizzes of length 9, 8, and so on. We then calculated the Spearman rank correlation between the image lists produced by \sys for each concept relative to the 10 question quiz. Figure~\ref {fig:asympt} shows these results. While we do not advocate for any specific number of questions, around a 6-question quiz we are just shy of a $.9$ correlation and any larger test exceeds that.

\section{Discussion}
\sys presents a novel approach to selecting the most visually appropriate image for Wikipedia articles using VLMs. Rather than relying on captions or prior indexing, our method evaluates images purely on their ability to help answer content-based visual questions. This acts as a proxy for the usefulness of the image, as a representative example of that concept. Contrastive question generation mechanism further enables disambiguation between visually similar candidates, an important step beyond surface-level matching. Our results provide evidence of the viability of the approach. However, it is worth considering how such a system would be implemented in practice. 

The problem of selecting images for Wikipedia articles shares characteristics with other editorial processes (e.g., picking images for textbooks, guidebooks, encyclopedias or dictionaries). However, there are also a number of unique constraints. Wikipedia images have specific intellectual property features (they need to be free) and it is likely that neither the editors or image creators (e.g., photographers or illustrators) are trained.  Additionally, not all topics have the same number of images, making some selection tasks easy and others much harder. For example, we see a range from the \textit{Nainamo bar's} (a dessert common in Nainamo, Canada) 15 images, to the \textit{Eastern Bluebird's} over 800, and the Indian rice beer \textit{Hariya} with none. This is due to in part to the unequal availability of images due to the concept's popularity~\citep{he2018the_tower_of_babel} and how contextually constrained it is (can someone take a picture of it?).

There are a number of opportunities to use \sys in other Wikipedia tools for editors. For example, rather than ranking images, \sys could also highlight gaps in existing images and encourage the collection or creation of new illustrative aids. Other extensions might take into account how questions could be weighted (e.g., through editor control or by looking at `longer-lived' Wikipedia edits about the way the concept looks). Future work may also include aesthetic features~\citep{shahid2014no}.

Our findings echo prior educational psychology research suggesting that well-chosen visuals can significantly enhance comprehension and engagement~\citep{carney2002pictorial}. In future work, we aim to extend our approach along several dimensions. First, we plan to generalize the framework to multilingual Wikipedia editions, which will require adapting both question generation and answer evaluation to respect language-specific descriptions, naming conventions, and cultural visual references. Second, we seek to enhance model robustness when dealing with culturally nuanced or historically sensitive concepts, where subtle visual cues or symbolic imagery may require careful treatment. To this end, incorporating human-in-the-loop validation and region-specific training data may help improve trustworthiness and fairness. Finally, we envision the development of interactive authoring tools that allow Wikipedia editors to collaborate with VLMs during image selection, offering question-based justifications, visual comparisons, and transparency into model decision-making, thus making AI-assisted curation both accessible and accountable.

\subsection{Limitations}
Although our approach demonstrates promising performance in selecting visually relevant images for Wikipedia articles, several limitations remain. First, our framework assumes that article content can be effectively translated into multiple-choice questions. This excludes concepts that are inherently abstract, historical, or biographical, where visual representation is unavailable or difficult to evaluate visually. Questions that focus on more than simple visual properties may be useful in expanding our approach. Second, our indexing control experiments provide some evidence of visual grounding, but stronger guarantees would require model transparency or architectural constraints. Third, the quality of generated questions is influenced by prompt engineering and concept length; very short or poorly written articles may yield sparse or low-quality visual questions. Finally, although we selected a variety of topic areas, our evaluation is limited to curated concepts in English Wikipedia. Future work should assess generalizability across languages, domains, and larger-scale deployments.

\section{Conclusions}
In this work, we introduce \sys, a system to rank images for use as illustrative aids. The central feature of our approach is that illustrative aids should be evaluated as educational interventions. By generating assessments directly from the content of a Wikipedia article, \sys can leverage a VLM to determine the effectiveness of the image in helping answer the questions. \sys offers additional features to generate contrastive questions that not only isolate good images for the class, but can highlight unique features of that class. Our evaluation determines the ability of VLMs to perform as a human would when viewing images and questions, and the effectiveness of \sys in identifying good images.

\bibliographystyle{aaai}
\bibliography{custom}

\begin{thebibliography}{}

\bibitem[\protect\citeauthoryear{Agrawal \bgroup et al\mbox.\egroup
  }{2011}]{agrawal2011enriching}
Agrawal, R.; Gollapudi, S.; Kannan, A.; and Kenthapadi, K.
\newblock 2011.
\newblock Enriching textbooks with images.
\newblock In {\em CIKM'11}.

\bibitem[\protect\citeauthoryear{Aletras and
  Stevenson}{2013}]{aletras2013representing}
Aletras, N., and Stevenson, M.
\newblock 2013.
\newblock Representing topics using images.
\newblock In {\em NAACL'13},  158--167.

\bibitem[\protect\citeauthoryear{Bruce and
  Humphreys}{1994}]{bruce1994recognizing}
Bruce, V., and Humphreys, G.~W.
\newblock 1994.
\newblock Recognizing objects and faces.
\newblock {\em Visual cognition} 1(2-3):141--180.

\bibitem[\protect\citeauthoryear{Carney and Levin}{2002}]{carney2002pictorial}
Carney, R.~N., and Levin, J.~R.
\newblock 2002.
\newblock Pictorial illustrations still improve students' learning from text.
\newblock {\em Educational psychology review} 14(1):5--26.

\bibitem[\protect\citeauthoryear{Clark and Lyons}{2010}]{clark2010graphics}
Clark, R.~C., and Lyons, C.
\newblock 2010.
\newblock {\em Graphics for learning: Proven guidelines for planning,
  designing, and evaluating visuals in training materials}.
\newblock John Wiley \& Sons.

\bibitem[\protect\citeauthoryear{Guo \bgroup et al\mbox.\egroup
  }{2020}]{guo2020you}
Guo, D.; Zhang, S.; Wright, K.~L.; and McTigue, E.~M.
\newblock 2020.
\newblock Do you get the picture? a meta-analysis of the effect of graphics on
  reading comprehension.
\newblock {\em AERA} 6(1).

\bibitem[\protect\citeauthoryear{Gyselinck and
  Tardieu}{1999}]{gyselinck1999role}
Gyselinck, V., and Tardieu, H.
\newblock 1999.
\newblock {\em The role of illustrations in text comprehension: What, when, for
  whom, and why?}
\newblock Mahwah, NJ, US: Lawrence Erlbaum Associates Publishers.
\newblock  195--218.

\bibitem[\protect\citeauthoryear{Haladyna}{2004}]{haladyna2004developing}
Haladyna, T.~M.
\newblock 2004.
\newblock {\em Developing and validating multiple-choice test items}.
\newblock Routledge.

\bibitem[\protect\citeauthoryear{Han \bgroup et al\mbox.\egroup
  }{2020}]{han2020extracting}
Han, S.; Ren, F.; Du, Q.; and Gui, D.
\newblock 2020.
\newblock Extracting representative images of tourist attractions from flickr
  by combining an improved cluster method and multiple deep learning models.
\newblock {\em Int. J. of Geo-Information} 9(2):81.

\bibitem[\protect\citeauthoryear{Harp and Mayer}{1997}]{harp1997role}
Harp, S.~F., and Mayer, R.~E.
\newblock 1997.
\newblock The role of interest in learning from scientific text and
  illustrations: On the distinction between emotional interest and cognitive
  interest.
\newblock {\em J. of educational psychology} 89(1):92.

\bibitem[\protect\citeauthoryear{He \bgroup et al\mbox.\egroup
  }{2018}]{he2018the_tower_of_babel}
He, S.; Lin, A.~Y.; Adar, E.; and Hecht, B.~J.
\newblock 2018.
\newblock The\_tower\_of\_babel. jpg: Diversity of visual encyclopedic
  knowledge across wikipedia language editions.
\newblock In {\em ICWSM'18}.

\bibitem[\protect\citeauthoryear{Kurdi \bgroup et al\mbox.\egroup
  }{2020}]{kurdi2020systematic}
Kurdi, G.; Leo, J.; Parsia, B.; Sattler, U.; and Al-Emari, S.
\newblock 2020.
\newblock A systematic review of automatic question generation for educational
  purposes.
\newblock {\em Int. J. of Artificial Intelligence in Education} 30(1):121--204.

\bibitem[\protect\citeauthoryear{Leake \bgroup et al\mbox.\egroup
  }{2020}]{leake2020generating}
Leake, M.; Shin, H.~V.; Kim, J.~O.; and Agrawala, M.
\newblock 2020.
\newblock Generating audio-visual slideshows from text articles using word
  concreteness.
\newblock In {\em CHI'20},  1--11.

\bibitem[\protect\citeauthoryear{Leggett and Kirchoff}{2011}]{plantguide}
Leggett, R., and Kirchoff, B.~K.
\newblock 2011.
\newblock {Image use in field guides and identification keys: review and
  recommendations}.
\newblock {\em AoB PLANTS} 2011.

\bibitem[\protect\citeauthoryear{Li \bgroup et al\mbox.\egroup
  }{2024}]{li2024generation}
Li, D.; Jiang, B.; Huang, L.; Beigi, A.; Zhao, C.; Tan, Z.; Bhattacharjee, A.;
  Jiang, Y.; Chen, C.; Wu, T.; et~al.
\newblock 2024.
\newblock From generation to judgment: Opportunities and challenges of
  llm-as-a-judge.
\newblock {\em arXiv preprint arXiv:2411.16594}.

\bibitem[\protect\citeauthoryear{Little \bgroup et al\mbox.\egroup
  }{2012}]{little2012multiple}
Little, J.~L.; Bjork, E.~L.; Bjork, R.~A.; and Angello, G.
\newblock 2012.
\newblock Multiple-choice tests exonerated, at least of some charges: Fostering
  test-induced learning and avoiding test-induced forgetting.
\newblock {\em Psychological science} 23(11):1337--1344.

\bibitem[\protect\citeauthoryear{Lu and Wang}{2024}]{lu2024generativestudents}
Lu, X., and Wang, X.
\newblock 2024.
\newblock Generative students: Using llm-simulated student profiles to support
  question item evaluation.
\newblock In {\em L@S'24},  16–27.
\newblock ACM.

\bibitem[\protect\citeauthoryear{Marsh and
  White}{2003}]{Marsh2003:0022-0418:647}
Marsh, E.~E., and White, M.~D.
\newblock 2003.
\newblock A taxonomy of relationships between images and text.
\newblock {\em J. of Documentation} 59(6):647--672.

\bibitem[\protect\citeauthoryear{Moore \bgroup et al\mbox.\egroup
  }{2023}]{moore2023assessing}
Moore, S.; Nguyen, H.~A.; Chen, T.; and Stamper, J.
\newblock 2023.
\newblock Assessing the quality of multiple-choice questions using gpt-4 and
  rule-based methods.
\newblock In {\em European Conf. on Technology Enhanced Learning},  229--245.
\newblock Springer.

\bibitem[\protect\citeauthoryear{Navarrete and
  Villaespesa}{2020}]{navarrete2020image}
Navarrete, T., and Villaespesa, E.
\newblock 2020.
\newblock Image-based information: paintings in wikipedia.
\newblock {\em J. of Documentation}.

\bibitem[\protect\citeauthoryear{Radford \bgroup et al\mbox.\egroup
  }{2021}]{radford2021learning}
Radford, A.; Kim, J.~W.; Hallacy, C.; Ramesh, A.; Goh, G.; Agarwal, S.; Sastry,
  G.; Askell, A.; Mishkin, P.; Clark, J.; et~al.
\newblock 2021.
\newblock Learning transferable visual models from natural language
  supervision.
\newblock In {\em ICML'21},  8748--8763.

\bibitem[\protect\citeauthoryear{Rama \bgroup et al\mbox.\egroup
  }{2022}]{rama2022large}
Rama, D.; Piccardi, T.; Redi, M.; and Schifanella, R.
\newblock 2022.
\newblock A large scale study of reader interactions with images on wikipedia.
\newblock {\em EPJ Data Science} 11(1):1.

\bibitem[\protect\citeauthoryear{Richards~Maldonado, Abouzied, and
  Gleason}{2023}]{richards2023readerquizzer}
Richards~Maldonado, L.; Abouzied, A.; and Gleason, N.~W.
\newblock 2023.
\newblock Readerquizzer: Augmenting research papers with just-in-time learning
  questions to facilitate deeper understanding.
\newblock In {\em Companion Pub. of CSCW'23},  391--394.

\bibitem[\protect\citeauthoryear{Rodriguez}{2005}]{https://doi.org/10.1111/j.1745-3992.2005.00006.x}
Rodriguez, M.~C.
\newblock 2005.
\newblock Three options are optimal for multiple-choice items: A meta-analysis
  of 80 years of research.
\newblock {\em Educational Measurement: Issues and Practice} 24(2):3--13.

\bibitem[\protect\citeauthoryear{Schwartz, Tsang, and
  Blair}{2016}]{schwartz2016abcs}
Schwartz, D.~L.; Tsang, J.~M.; and Blair, K.~P.
\newblock 2016.
\newblock {\em The ABCs of how we learn}.
\newblock WW Norton \& Company.

\bibitem[\protect\citeauthoryear{Shahid \bgroup et al\mbox.\egroup
  }{2014}]{shahid2014no}
Shahid, M.; Rossholm, A.; L{\"o}vstr{\"o}m, B.; and Zepernick, H.-J.
\newblock 2014.
\newblock No-reference image and video quality assessment: a classification and
  review of recent approaches.
\newblock {\em J. on Image and Video Processing} 2014(1):40.

\bibitem[\protect\citeauthoryear{Shimmei, Bier, and
  Matsuda}{2023}]{shimmei2023machine}
Shimmei, M.; Bier, N.; and Matsuda, N.
\newblock 2023.
\newblock Machine-generated questions attract instructors when acquainted with
  learning objectives.
\newblock In {\em Int. Conf. on Artificial Intelligence in Education},  3--15.
\newblock Springer.

\bibitem[\protect\citeauthoryear{Silva \bgroup et al\mbox.\egroup
  }{2024}]{silva2024imagine}
Silva, A.; Tracy, M.; Reinecke, K.; Adar, E.; and Redi, M.
\newblock 2024.
\newblock Imagine a dragon made of seaweed: How images enhance learning in
  wikipedia.
\newblock {\em arXiv preprint arXiv:2403.07613}.

\bibitem[\protect\citeauthoryear{Singh, Zouhar, and
  Sachan}{2023}]{singh2023enhancing}
Singh, J.; Zouhar, V.; and Sachan, M.
\newblock 2023.
\newblock Enhancing textbooks with visuals from the web for improved learning.
\newblock {\em arXiv preprint arXiv:2304.08931}.

\bibitem[\protect\citeauthoryear{Srinivasan \bgroup et al\mbox.\egroup
  }{2021}]{srinivasan2021wit}
Srinivasan, K.; Raman, K.; Chen, J.; Bendersky, M.; and Najork, M.
\newblock 2021.
\newblock Wit: Wikipedia-based image text dataset for multimodal multilingual
  machine learning.
\newblock {\em arXiv preprint arXiv:2103.01913}.

\bibitem[\protect\citeauthoryear{Stein}{1991}]{stein1991illustrations}
Stein, G.
\newblock 1991.
\newblock Illustrations in dictionaries.
\newblock {\em Int. J. of lexicography} 4(2):99--127.

\bibitem[\protect\citeauthoryear{Sutton}{2021}]{birdguide}
Sutton, C.
\newblock 2021.
\newblock Can photo guides do it all?
\newblock \url{https://www.aba.org/can-photo-guides-do-it-all/}.

\bibitem[\protect\citeauthoryear{Wei \bgroup et al\mbox.\egroup
  }{2022}]{wei2022chain}
Wei, J.; Wang, X.; Schuurmans, D.; Bosma, M.; Xia, F.; Chi, E.; Le, Q.~V.;
  Zhou, D.; et~al.
\newblock 2022.
\newblock Chain-of-thought prompting elicits reasoning in large language
  models.
\newblock {\em Advances in neural information processing systems}
  35:24824--24837.

\bibitem[\protect\citeauthoryear{{Wikimedia}}{2022}]{wikimedia2022addimage}
{Wikimedia}.
\newblock 2022.
\newblock Add-an-image structured task.
\newblock \url{https://w.wiki/4LHR}.

\bibitem[\protect\citeauthoryear{Wikimedia}{2025}]{wikimg}
Wikimedia.
\newblock 2025.
\newblock Wikipedia:manual of style/images.
\newblock \url{https://w.wiki/6HT7}.

\bibitem[\protect\citeauthoryear{Yang and Lin}{2024}]{yang2024toward}
Yang, J.-H., and Lin, J.
\newblock 2024.
\newblock Toward automatic relevance judgment using vision--language models for
  image--text retrieval evaluation.
\newblock {\em arXiv preprint arXiv:2408.01363}.

\end{thebibliography}

\appendix

\section{Appendix A: Prompts}
\subsection{Question Generation}\label{sec:mcqprompt}
\begin{lstlisting}[basicstyle=\footnotesize]
I will give you an article about something. I would like to generate a multiple choice test to give to people that they should be able to answer if they've seen (or a picture of it). Generate as many questions as you can. You will do this in two steps. First, reason about what someone should be able to answer if they have visually seen it. Second, generate questions based on this reasoning. Specifically, describe how an image could be used to answer the question. In addition, generate questions that focus on visually identifiable features only, such as physical attributes, textures, presentation style, or objects present in the image. Avoid questions about specific locations, measurements, or other details that cannot be directly observed or inferred from the image alone. Then, the questions cannot be too easy. For the generated questions, do not mention the specific object (directly ask the name of object). For example, How would you describe the meat in the image?, How is the object in the image typically served?
   
Generate multiple-choice questions focusing on special features of the object according to the text that the other similar objects do not have. Avoid generating questions about very common features. The format of the generated questions should be JSON as follows.
The format is:
    [
     {
        "question": "Your question here",
        "options": ["A) Option 1", "B) Option 2", "C) Option 3", "D) Option 4"],
        "correct_answer": "Correct Option",
        "rationale": "Explanation here"
     },
    ...
    ]
\end{lstlisting}

\subsection{Comparative Question Generation}\label{sec:compprompt}
The following is an example prompt for generating questions that factor in the visual distinctiveness of the target concept A versus B.

\begin{lstlisting}[basicstyle=\footnotesize]
Analyze the descriptions of Object A and Object B below and identify their distinct differences in visual appearance.  Focus on the unique visual features of Object A that distinguish it from Object B, and vice versa.  Do not include features based on abstract or non-visual attributes such as purpose, measurement, utility, or contextual associations. Focus exclusively on observable physical characteristics, such as shape, color, texture, size, patterns, or structural details. 
1. Identify and list the distinct visual features of A and B in the following format:
   (set 1) Features distinct to A:
       A. [Feature A]
       B. [Feature B]
       C. [Feature C]
       ...
   (set 2) Features distinct to B:
       D. [Feature D]
       E. [Feature E]
       F. [Feature F]
       ...
   (set 3) Features that both A and B have in common:
       G. [Feature G]
       H. [Feature H]
       I. [Feature I]
       ...
2. Based on the identified features, generate multiple-choice questions where:
   - The correct answer is only applicable to Object A (from set 1).
   - The distractors are features that are applicable only to Object B (from set 2),and the correct answer is randomly among A,B,C,D.
   - Do not generate the repeated questions or choices in set of questions that you generate,and use a variety of question formats.
   - Do not generate comparable questions, like how different with others.
   - Do not generate the choices which are directly conflicting with questions. In other word, we need to exclude distractors by images rather than the choices are not related to questions distinctly. 
   - For the generated questions,do not mention the specific object.
   - Do not include the answer directly in the question. Instead, ask indirect questions that require the user to infer the correct answer.
   - Use a variety of **generic question formats** that are applicable to any topic, such as:
     - 'What is a key feature of the object in the image?'
     - 'What characteristic is presented in the image?'
     - 'What is visible in the image?'
     - 'What feature is depicted in the image?'
     - 'What is a key aspect of the image?'
   - Structure the questions in the following JSON format:
     [
       {
         "question": "Your question here",
         "options": ["A) Option 1", "B) Option 2", "C) Option 3", "D) Option 4"],
         "correct_answer": "Correct Option",
         "rationale": "Explanation of why the answer is correct."
       },
       ...
     ]
\end{lstlisting}

\clearpage
\subsection{VLM Instructions}\label{sec:vlmprompt}
\begin{lstlisting}[basicstyle=\footnotesize]
You must answer this question based on the visible information in the image, and there is only one correct answer for every question. 
Do not use prior knowledge, indexed knowledge beyond what is observed in the image. 
If the required information is not visible in the image or you think there is no correct option given, you should answer with: 'I can't answer that based on the image.'.
Use the chain of thought to answer all questions including the questions you can not answer, which means that give the reason why you can not answer. 
Your output must be in the format:
Analysis: <your reasoning>\n
Final answer: <your selected answer or 'I can't answer that based on the image.'>.
\end{lstlisting}

\section{Appendix B: Case Study}
\label{appendix:gujia_case_study}

This appendix presents a full case study of how the model performs on two visually similar food items: \textit{Gujia} (target) and \textit{Chandrakala} (distractor).

\begin{figure}[ht]
    \centering
    \includegraphics[width=0.4\textwidth]{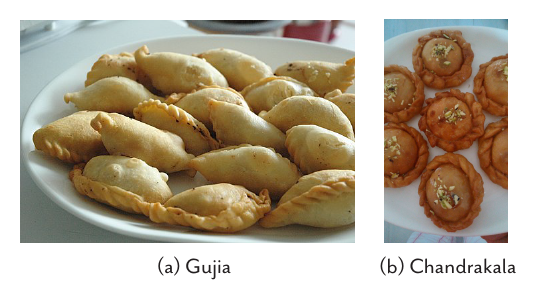}
    \caption{(a) An image of the target concept (gujia) and (b) a visually similar distractor (chandrakala).}
    \label{fig:gujia_chandrakala_imgs}
\end{figure}

\begin{table}[ht]
\centering
\scriptsize
\renewcommand{\arraystretch}{1.3}
\begin{tabularx}{\linewidth}{>{\raggedright\arraybackslash}X c c}
\toprule
\textbf{Original Question} & \textbf{Gujia} & \textbf{Chandrakala} \\
\midrule
What distinct shape does the sweet dumpling have? & correct (Half moon) & incorrect (Round) \\
How would you describe the texture of the outer layer? & incorrect (Smooth) & incorrect (Smooth) \\
What is the typical color when fully prepared? & correct (Golden-brown) & correct (Golden-brown) \\
Ingredient likely in the filling? & incorrect (No answer) & correct (Khoa) \\
Which element is least likely a visual feature? & correct (Icing) & incorrect (Half moon) \\
\midrule
\textbf{Total Correct} & \textbf{3/5} & \textbf{2/5} \\
\bottomrule
\end{tabularx}
\caption{Model performance on original questions. Both images are answered similarly (discriminative failure).}
\label{tab:original_q_result}
\end{table}

\begin{table}[ht]
\centering
\scriptsize
\renewcommand{\arraystretch}{1.4}
\begin{tabularx}{\linewidth}{
  >{\raggedright\arraybackslash}p{0.36\linewidth} 
  >{\raggedright\arraybackslash}p{0.28\linewidth} 
  >{\raggedright\arraybackslash}p{0.28\linewidth}
}
\toprule
\textbf{Contrastive Question} & \textbf{Gujia} & \textbf{Chandrakala} \\
\midrule
What is a key feature of the object in the image? & correct (Half-moon) & incorrect (Sun-shaped) \\
Which of the following is unique to the object shown? & correct (Empanada-like) & incorrect (Moonlight) \\
How does the object differ from others? & correct (Single fold) & incorrect (Two circles) \\
What is visible in the image? & correct (Empanada shape) & incorrect (Rounded dough circle) \\
\midrule
\textbf{Total Correct} & \textbf{4/4} & \textbf{0/4} \\
\bottomrule
\end{tabularx}
\caption{Model performance on contrastive questions.}
\label{tab:contrastive_q_result}
\end{table}

\subsection*{Detailed Model Reasoning (Contrastive Questions)}
\begin{footnotesize}
\paragraph{Q1: What is a key feature of the object in the image?}  
\textbf{Correct Answer:} B) Half-moon shaped

\textit{Gujia:}  
"The objects in the image have a folded appearance with crimped edges, resembling a crescent or half-moon shape. This suggests the best option is: B) Half-moon shaped."  
\textbf{Correct}

\textit{Chandrakala:}  
"The objects appear rounded with a decorative edge. Based on the sun-like appearance, the model selects: C) Sun-shaped using two dough circles."  
\textbf{Incorrect}

\paragraph{Q2: Which of the following is unique to the object shown?}  
\textbf{Correct Answer:} B) Looks like an empanada

\textit{Gujia:}  
"The appearance closely resembles an empanada, with a half-moon shape and crimped edge. Best answer: B) Looks like an empanada."  
\textbf{Correct}

\textit{Chandrakala:}  
"The object seems more circular and enclosed. Based on that, selects: C) Associated with moonlight shape."  
\textbf{Incorrect}

\paragraph{Q3: How does the object differ from others?}  
\textbf{Correct Answer:} B) Uses a single layer of dough folded over

\textit{Gujia:}  
"The crescent shape suggests they are folded over from a single layer. Correct choice: B)."  
\textbf{Correct}

\textit{Chandrakala:}  
"Flat-topped and round, suggesting two separate dough layers. Selects: D) Two circles for sun shape."  
\textbf{Incorrect}

\paragraph{Q4: What is visible in the image?}  
\textbf{Correct Answer:} B) A shape that mimics an empanada

\textit{Gujia:}  
"Classic empanada-like semicircular shape with crimped edges. Selects: B)."  
\textbf{Correct}

\textit{Chandrakala:}  
"Dome-like and circular, with no folding. Selects: C) Rounded from a single circle of dough."  
\textbf{Incorrect}

\subsection*{Conclusion}

This case demonstrates that original questions resulted in a narrow performance gap (Gujia 3/5, Chandrakala 2/5), while contrastive questions achieved clear separation (Gujia 4/4, Chandrakala 0/4).   
\end{footnotesize}

\end{document}